\documentclass[useAMS,usenatbib]{mn2e}

\usepackage{amsmath}
\usepackage{graphicx}
\usepackage{color}
\usepackage{ulem}
\usepackage{amssymb}
\usepackage{bm}
\usepackage{hyperref}
\usepackage[dvipsnames]{xcolor}

\usepackage{epsf}
\def\msun{{\rm M_{\odot}}}

\def\me{{\dot M_{\rm Edd}}}
\def\le{{L_{\rm Edd}}}

\title[No magnetars in ULXs] {No magnetars in ULXs}

\author[Andrew King \& Jean--Pierre Lasota]
{Andrew King$^{1, 2, 3, 4}$ \& Jean--Pierre Lasota$^{4, 5}$\\
$^{1}$ Theoretical Astrophysics Group, Department of Physics \& Astronomy, University of Leicester, Leicester LE1 7RH, UK\\
$^{2}$ Astronomical Institute Anton Pannekoek, University of Amsterdam, Science Park 904, 1098 XH Amsterdam, Netherlands\\
$^{3}$ Leiden Observatory, Leiden University, Niels Bohrweg 2, NL-2333 CA Leiden, Netherlands\\
$^{4}$ Institut d'Astrophysique de Paris, CNRS et Sorbonne Universit\'e, UMR 7095, 98bis Bd Arago, 75014 Paris, France\\  
$^{5}$ Nicolaus Copernicus Astronomical Center, Polish Academy of Sciences, ul. Bartycka 18, 00-716 Warsaw, Poland\\     
}

\date{\today}

\volume{000}

\pagerange{\pageref{firstpage}--\pageref{lastpage}} \pubyear{2019}

\begin{document}

\label{firstpage}

\maketitle

\begin{abstract}
We consider the current observed ensemble of pulsing ultraluminous X--ray sources (PULXs). We
show that all of their observed properties (luminosity, spin period, and spinup rate)
are consistent with emission from magnetic neutron stars with 
fields in the usual range $10^{11} - 10^{13}\, {\rm G}$, which is collimated (`beamed')  by
the outflow from an accretion disc supplied with mass at a super--Eddington rate, but ejecting
the excess, in the way familiar for other (non--pulsing) ULXs. The observed properties are
inconsistent with magnetar--strength fields in all cases. 
We point out that all proposed pictures of magnetar formation suggest that they are
unlikely to be members of binary systems, in agreement with the observation that all confirmed 
magnetars are single. The presence of magnetars in ULXs is therefore 
improbable, in line with our
conclusions above.
\end{abstract}


\section{Introduction}

Ultraluminous X--ray sources (ULXs) are defined by apparent (assumed isotropic)  luminosities 
$L\ga 10^{39}\, {\rm erg\, s^{-1}}$, above the usual Eddington 
value for stellar--mass objects, but which do not contain supermassive black holes. When they
were first identified around the turn of the century, the natural assumption was to regard these 
apparent luminosities as intrinsic, which then required that   
their accretors should be black holes with masses intermediate
between stellar and supermassive \citep[e.g.,][]{CM99}. 

But by now it is generally
accepted \citep[cf][]{Kaaret17} that  most (or possibly all) ULXs are 
are otherwise normal stellar--mass X--ray binaries in an unusual evolutionary state.
Their intrinsic luminosities do not significantly exceed Eddington, but appear so when seen 
in a narrow range of viewing angles: they are not isotropic but collimated (or `beamed')
by some factor $b < 1$. Then (incorrectly) assuming isotropic emission suggests a larger luminosity
\begin{equation}
L_{\rm sph} = \frac{1}{b}L
\label{eq:lbeam}
\end{equation}
than the correct value $L \simeq \le$, where
\begin{align}
L_{\rm Edd} & =\frac{4\pi c GM}{\kappa}\nonumber \\
& =2.5\times 10^{38}(1+X)^{-1} m_1\,\rm erg\,s^{-1}
\label{eq:Ledd}
\end{align}
is the Eddington luminosity. In Eq. (\ref{eq:Ledd})  $\kappa=\sigma_T/m_p$, where 
$\sigma_T$ is the Thompson
scattering cross-section and $m_p$ is the proton mass; $c$ is the speed of light. 
$m_1\equiv {M}/{\msun}$, is the accretor mass in solar units, and $X$ the hydrogen abundance 
by mass.

Perhaps confusingly, the reason for the beaming is that mass {\it transfer} rates in ULX
binaries are highly super--Eddington (defining the unusual evolutionary state) 
but mass {\it accretion} is effectively only Eddington.
As envisaged by \citet{SS73} the excess is ejected in a quasi--spherical outflow whose
collimating structure is the cause
of the anisotropic luminosity.

Strong support for this picture comes from the 
fact that at least one neutron star, that in the low--mass X--ray binary Cygnus X--2, has survived
being fed $\sim 3\msun$ from its (previously more massive) companion star at highly
super--Eddington rates ($\sim 10^{-5}\msun\, {\rm yr}^{-1}$), yet has evidently gained no more 
than $\sim {\rm few}\times 0.2\msun$ \citep{KR99}, apparently 
ejecting all the surplus.

This agrees with the reasoning of \citet{SS73} concerning accretion discs fed at super--Eddington rates.  Mass is progressively blown out of the disc 
near the accretor so that the local Eddington limit is respected at each disc radius $R$, making the 
local accretion rate $\dot M(R)$ decrease linearly with $R$. This
outflow leaves only narrow channels around the rotational axis of the
accretion disc for the accretion luminosity to escape, giving an effective beaming factor $b <1$
\citep[cf][]{King09}. The extreme Galactic accreting system SS433 also has this structure, 
but we do not view it along
the beam axis \citep[][]{Begelman06,vdH17} so it is a ULX
seen `from the side'. SS433 was found because of its unusual precessing near--relativistic jets, but
in general we cannot easily detect ULX systems where we are not on the beam axis \citep[see][]{MiddletonSS}.

As \citet{Kingetal01} pointed out, because they result from beaming,
ULX luminosities alone do not directly tell us the nature of the
accretor. Although the initial tendency was to assume that the accretors were all black holes,
they may instead be neutron stars or even white dwarfs, provided that
their mass transfer rates are $\ga 10^{-8}, 10^{-5}\msun\,{\rm  yr}^{-1}$ respectively. 
(In principle rates $\ga 3\times 10^{-4}\msun\,{\rm  yr}^{-1}$ would make
main--sequence accretors super--Eddington, but X--ray production is less likely.)
Following this prediction \citet{Fabbianoetal03} suggested that one of the ULXs in the 
Antennae probably involves an accreting white dwarf because of its unusually soft emission. 

The advent of NuSTAR allowed a further confirmation of this prediction, a ULX (M82 XÐ2)
with a coherent periodicity P = 1.37 s, naturally interpreted as
the spin period of an accreting magnetic neutron star \citep{Bachettietal14}. 
Its apparent luminosity
$L \simeq 1.8\times 10^{40}\, {\rm erg\, s}^{-1}$ is about 100 times Eddington.  
Because electron--scattering cross sections are reduced for some directions and
polarizations for very strong--field systems such as magnetars, this raised the possibility that 
the apparent super--Eddington luminosity might be intrinsic \citep[cf][]{Tong15, Eksi15,Dosso15}. In this sense the idea of magnetar--strength fields for PULXs is a magnetic analogue of the 
suggestion of intermediate--mass black holes for non--pulsing ULXs.  \citet{KL15} already
pointed out that this idea led to problems because the accretion disc would be disrupted very
far from the neutron star. This would imply a huge accretion lever--arm and a spinup
rate far larger than observed. Using the observed rate, \citet{KL15} instead
postulated a magnetic fieldstrength typical of a millisecond pulsar.

\begin{figure}
\begin{center}
\includegraphics[width=\columnwidth]{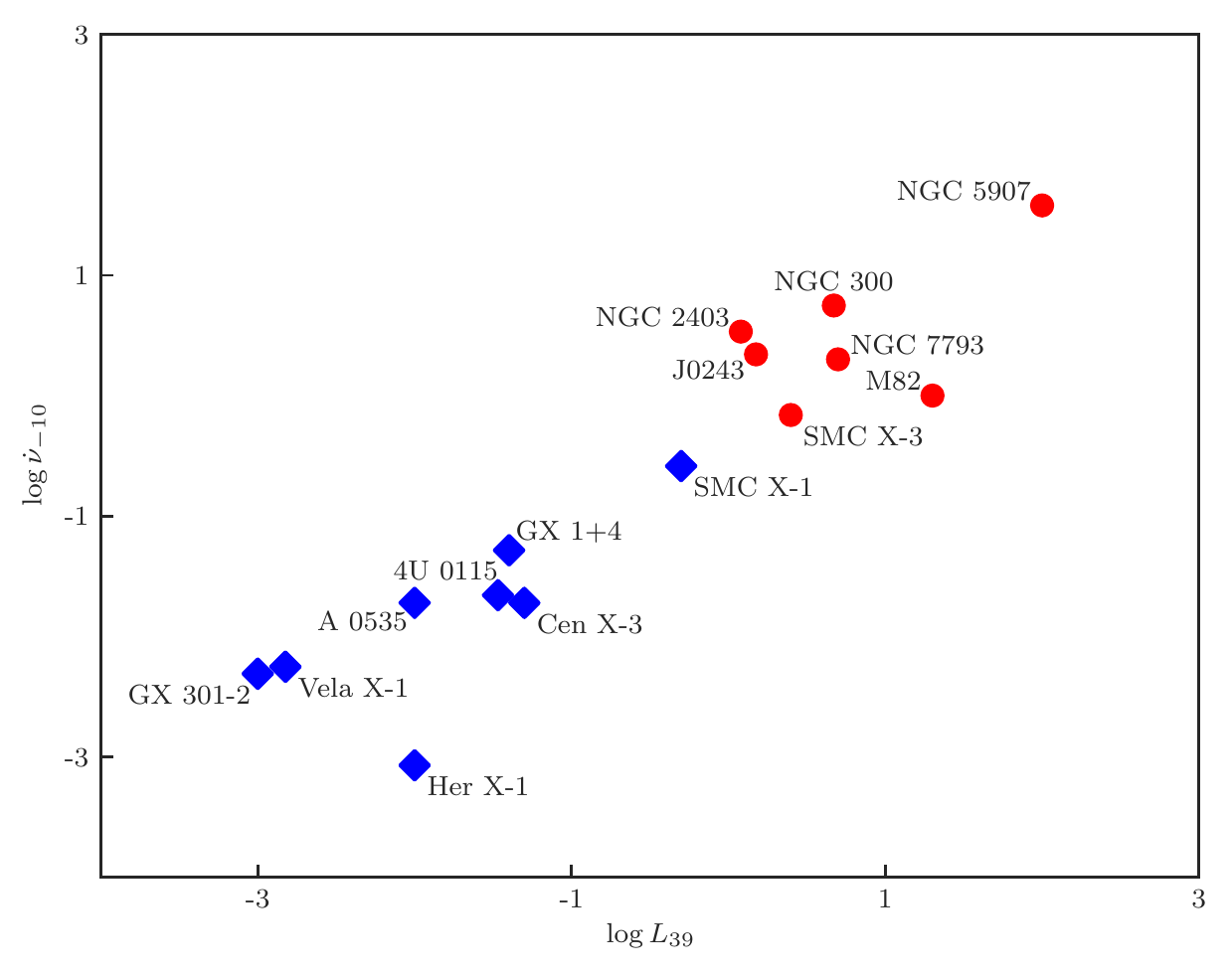} 
\caption{The $L_{39}$ -- $\dot \nu_{-10}$ diagram for XRPs and PULXs. Red dots: the seven PULXs with known spin-up rates. Blue diamonds:
selected (for comparison) X-ray pulsars (see Table in the Appendix)}
\label{fig:dotnul}
\end{center}
\end{figure}

In line with this, \citet{KL16} showed that M82 X--2 
fitted naturally into the simple picture of ULXs as beamed X--ray sources, fed at
super--Eddington mass transfer rates but accreting only about Eddington, as already established for 
unpulsed ULXs. It followed that
its magnetic field should indeed be weaker ($\simeq 10^{11}$\,G) than in a
young X--ray pulsar, as expected if the neutron star had gained mass. This unified picture got further
impetus from a characteristic property of the beaming formula
\begin{equation}
b \simeq \frac{73}{\dot m^2}
\label{eq:b}
\end{equation}
suggested by \citet{King09} on the basis of a combination of theoretical arguments and
observations
($\dot m = \dot M/\me$ is the accretor's Eddington factor, where $\dot M$ is the mass 
transfer rate). This
implies that for a given $\dot M$, a neutron star accretor has a {\it higher} apparent luminosity
than a more massive black hole. Using 
$\dot m \propto M^{-1}$ and  $\le \propto M$,  equations (\ref{eq:lbeam}, \ref{eq:b})
give $L_{\rm sph} \propto M^{-1}$. Accordingly, \citep[][hereafter KLK17]{PULX} suggested
that a significant fraction of non--pulsing 
ULXs might actually contain neutron stars rather than black 
holes \citep[see also][]{Koliopanos17}. KLK17 analysed the three then known pulsing ULXs (PULXs), adopting super--Eddington 
mass transfer as the defining characteristic of ULXs. They showed that all three systems
had magnetospheric radii $R_M$ very close to the spherization radii 
$R_{\rm sph}$ where radiation pressure becomes important and drives mass
loss from the accretion disc. KLK17 argued that the condition
\begin{equation}
R_M \sim R_{\rm sph}
\label{eq:rm-rsph}
\end{equation}
is very probably necessary
for pulsing. $R_{\rm sph}$ is only defined at all if it is larger than $R_M$, but if
$R_{\rm sph} \gg R_M$ the pulse fraction must be small, so the system would probably appear
as an unpulsed ULX.
KLK17 also showed that (\ref{eq:rm-rsph}) requires
PULX spinup rates 
\begin{equation}
- \frac{\dot P}{P^2}= \dot\nu > 10^{-10}\, {\rm s^{-2}}, 
\label{dotnu}
\end{equation}
more than 10 
times larger than any other pulsing neutron star (see Fig. \ref{fig:dotnul}). This in turn probably means that, if the system 
stays at super-critical luminosities, the spin period 
oscillates very rapidly around its `equilibrium' value as the accretion rate varies, with spinup
episodes easier to observe than spindown ones, where presumably accreting matter is centrifugally
repelled. Searches for propeller--phase ULXs have not been so far successful \citep{Earnshow18}.
In the three systems analysed by KLK17 the magnetic field
is self--consistently in the usual neutron--star range $10^{11} - 10^{13}\, {\rm G}$, again arguing 
against a magnetar origin for ULX behaviour.

Since pulsing is hard to see if $R_{\rm sph} \gg R_M$,
KLK17 suggest that the fraction of PULXs observed among
ULXs may severely underestimate the true fraction of neutron star ULXs.
Recent confirmation of this 
comes from the discovery that one non--pulsing ULX \citep{Brightman18}
probably has a magnetized neutron--star accretor, revealed by a cyclotron 
line in its X--ray spectrum. This was initially identified as a highly unusual {\it proton} line corresponding
to $B \sim 10^{15} \, {\rm G}$ \citep{Brightman18}, but
recent re--analysis instead suggests an electron cyclotron line with
a more normal (pulsar--like) field of $B \sim 10^{11}- 10^{12}\, {\rm G}$ \citep{Middleton18}.  
Another cyclotron line corresponding to a $\sim 10^{12}$G field is seen in the PULX NGC300 ULX1
\citep[][see, however, \cite{Koliopanos18}]{Walton18}.

Theoretically the NS/BH fraction in ULXs is 
uncertain: evolutionary calculations 
show that it is likely to depend on the star formation history of the host galaxies \citep{Wiktor17,Wiktor18}. So determining the true fraction could have major implications
for understanding this history.

Observations relevant to this problem have made rapid progress since the analysis by
KLK17. There are now nine identified neutron--star ULXs 
(NSULXs; see Table \ref{tab:ulx1bis}), of which eight are observed as 
PULXs. Four of the NSULXs (M82 ULX2, NGC7793 P13, NGC5907 ULX1 and M51 ULX8) are 
`standard' ULXs, where the mass transfer rate seems to be fairly stably super--Eddington.
The observed variability of these bright ULXs might result from precession \citep[e.g.,][]
{Motchetal14} -- for example
an accretion disc irradiated by its central source may warp and precess \citep[][]{Pringle96,OG01}.
The correlation between the X--ray luminosity $L_X$ and spectral hardness, 
as well as changes in low resolution atomic absorption lines -- \cite{Middleton15} -- 
appear to support this idea.

The five other PULXs (including one in the Galaxy) are transient, and probably Be--X--ray binaries. 
Here a neutron star
is in an eccentric orbit around a Be star, accreting from its circumstellar disc at periastron in 
a quasi--periodic way. At intervals of $\sim 10$ orbits, a much brighter accretion episode occurs,
probably because of the Kozai--Lidov effect of the misaligned neutron--star orbit on the
Be star disc \citep{Martinetal14}. The system becomes a ULX 
only in these bright outburst phases. The pulsing X--ray sources in Be--ULXs are visible during the 
sub--Eddington phase. 
The increase of luminosity by 2--4 orders of magnitude does not suppress pulsations but does
increase $\dot \nu$ by at least an order of magnitude. In terms of Eq. \ref{eq:rm-rsph}, 
these `giant outbursts' presumably supply mass at super--Eddington rates, so that
$R_{\rm sph}$ is defined but only barely $\gtrsim R_M$. 
All Be--ULXs are relatively faint ($L_X < 4 \times 10^{39}$\,erg\,s$^{-1}$) compared with other
ULXs.

\begin{center}
\begin{table*}
{
\setlength{\tabcolsep}{1pt}
\caption{Observed properties of NSULXs}
\label{tab:ulx1bis}
{\small
\hfill{}
\begin{tabular}{ ||l||c||c||c||c||c|} 
 \hline\hline
  Name & $L_X (\rm max)$  [erg\,s$^{-1}$] & $P_s $ [s] &  $\dot \nu$ [s$^{-2}$] & $P_{\rm orb}$ [d]& $M_2$ [$\rm M_{\odot}$] \\
  \hline\hline
  M82 ULX2$^1$ & $2.0 \times 10^{40}$  & 1.37 \ \ & $ 10^{-10}$ \ \ &   2.51 (?) &  $\ga 5.2$ \\
   \hline
  NGC7793 P13$^2$  & $5\times 10^{39}$ & 0.42  \ \  &  $2 \times 10^{-10}$ \ \  &  63.9 & 18--23 (B9I) \\
   \hline
  NGC5907 ULX1$^3$ &  $\sim 10^{41}$ & 1.13 \ \ & $3.8 \times 10^{-9}$ \ \  & 5.3(?) &  \\
   \hline
  NGC300 ULX1$^4$ & $4.7\times 10^{39}$&  $\sim$31.5 \ \  &  $5.6 \times 10^{-10}$ \ \ &  $> 8$   & 40 (Be ?) \\
   \hline
  SMC X-3$^{5,6}$&  $ 2.5 \times 10^{39}$& $\sim 7.7$ \ \  & $6.9 \times 10^{-11}$ \ \ & 45.04 & $> 3.7 $(Be ?) \\
   \hline
  NGC 2403 ULX$^{7}$ & $1.2 \times 10^{39} $ &$\sim 18$ \ \ &   $3.4 \times 10^{-10}$ \ \ & 60 -- 100 (?)& (Be ?)\\
   \hline
  Swift J0243.6+6124$^{8}$ &  $ \gtrsim 1.5 \times 10^{39}$ (?) &  9.86  &  $2.2. \times 10^{-10}$ & 28.3 &  (Be ?)  \\
  \hline   
  NGC 1313 PULX$^9$ & $1.6 \times 10^{39} $ & $\sim 765.6$ \ \ &  & & (Be ?) \\
  \hline
  M51 ULX8$^10$  & $2\times 10^{39}$  & \rm NO & \rm NO &  8 -- 400 (?) &  40 (?) \\                                                                                                          
   \hline\hline
\end{tabular}}
}
\hfill{}
\vskip 0.2truecm 
$^1$\citet{Bachettietal14}, $^2$\citet{Furstetal16,Furstetal18,Israeletal17,Motchetal14}; ,\\ $^3$\citet{Israeletal16} 
$^4$\citet{Carpanoetal18}, $^5$\citet{2017A&A...605A..39T}, $^6$\citet{Townsend17}, $^7$\citet{2007ApJ...663..487T}, $^{8}$\citet{Doroshenko18}, $^9$\citet{Trudolyubov08}, $^{10}$\citet{Brightman18}.
\vskip 0.2truecm
\end{table*}
\end{center}

We have three aims in this paper. First, we extend the analysis of KLK17
to the rather larger current sample of ULXs to see if the condition (\ref{eq:rm-rsph})
is required for pulsing and (\ref{dotnu}) holds, and see if the field deduced from the cyclotron
line agrees with this analysis.  Second, we use this analysis to check whether
magnetar--strength fields are required in PULXs, finding that they are not, and indeed 
are inconsistent with the observations. Finally, considering formation mechanisms for
magnetars, we show that their presence in \textsl{any binary system} at all appears unlikely.

\begin{table*}
{
\setlength{\tabcolsep}{1pt}
\caption{Derived properties of PULXs}
\label{tab:ulx3b}
{\small
\hfill{}
\begin{tabular}{ ||l|||c||c||c||c||c||c||c||} 
 \hline\hline
 Name &  $\dot m_0$ \ \ &  ${\bm \mu}\, q^{7/4}m_1^{-1/2}I_{45}^{-3/2}$ [Gcm$^3$]& ${\bm R_{\rm sph}}m_1^{-1}$ [cm] &  ${\bm R_M} m_1^{-1/3}I_{45}^{-2/3}$ [cm] & ${\bm R_{\rm co}}m_1^{-1/3}$[cm] & ${\bm P_{\rm eq}}q^{-7/6}m_1^{1/3}$ [s] &  ${\bm t_{\rm eq}}$ [yr]$^1$\\
 \hline\hline
 M82 ULX2 &   36  & $9.0\times 10^{28}$ & $3.6\times 10^7$ & $1.0\times 10^7$ & $1.9\times 10^8$ &  0.02 & 15600 \\
 \hline
 NGC 7793 P13  &  20 &  $2.5\times 10^{29}$ & $2.1\times 10^7$ & $1.6\times 10^7$& $8.4 \times 10^8$ &  0.09 & 1386 \\
 \hline
 NGC5907 ULX1 & 91 & $2.1\times 10^{31}$ & $9.1\times 10^7$ & $1.1\times 10^8 $ & $1.6 \times 10^8$ & 1.86 &  0 \\
 \hline
 NGC300 ULX1 & 20 & $1.2\times 10^{30}$$^{\heartsuit}$ & $2.1\times 10^7$ &  $3.2 \times 10^7$& $1.5\times 10^9$ & 0.19& 297 \\
 \hline
 SMC X-3& 18 & $2.3\times 10^{28}$ & $1.8 \times 10^7$ &  $7.1 \times 10^6$& $5.9 \times 10^8$& 0.006&76621 \\
 \hline
 NGC 2403 ULX & 11 & $5.6\times 10^{29}$ & $1.1\times 10^7$ & $2.3 \times 10^7$ & $1.1\times 10^9$ & 0.16& 578 \\
 \hline
 Swift J0243.6+6124 & 14  & $1.6\times 10^{29}$  &  $1.4\times 10^7$ & $1.7\times 10^7$ & $6.9\times 10^8$  & 0.07 & 2047 \\       
 \hline
 NGC 1313 ULX & 14 & & $ 2.8 \times 10^7$&  & $8.8 \times 10^{13}$ & & \\
 \hline
 M51 ULX8& 16 & $\sim 3\times  10^{29}$$^{\clubsuit}$ & $1.6\times 10^7$ &  $2.7 \times 10^7$$^{\spadesuit}$ & ? & ?\\
\hline\hline.
\end{tabular}}
}
\hfill{}
\vskip 0.2truecm 
{$^1$ - calculated using the value of  ${\bm P_{\rm eq}}q^{-7/6}m_1^{1/3}$ from the previous column.}\\
$^{\heartsuit}$-- close to $B\sim 10^{12}$G as measured by \citet{Walton18}.\\
$^{\clubsuit}$-- from observations  \citep{Brightman18,Middleton18}.
$^{\spadesuit}$-- from Eq. (\ref{eq:rm}) for a $10^{11-12}$\,G magnetic field.
\end{table*}

\section{Properties of Neutron Star ULXs}

\subsection{``Standard" NSULXs}

\subsubsection{M82 ULX2}
ULX2 in the galaxy M82, also known as NuSTAR J095551+6940.8, was the first ultraluminous 
X--ray source
found to be an X--ray pulsar \citep{Bachettietal14}, ending the controversy about the presence 
of stellar-mass compact objects in ULXs and confirming the predictions of \citet{Kingetal01}.

\subsubsection{NGC7793 P13}
P13 in the galaxy NGC7793, also known as XMMU J235751.1-323725, was identified as an 
X--ray pulsar by \citet{Furstetal16} and \citet{Israeletal17}. \citet{Motchetal14} had already 
shown that the compact object in this system has a mass $\lesssim 15\msun$.
It is the only ULX with a well-identified companion: a BI9 supergiant \citep{Motch11}. The 
63.9 d binary orbit has eccentricity $e \leq 0.15$ \citep{Furstetal18}.

\subsubsection{NGC5907 ULX1}
\citet{Israeletal16} found that 
ULX1 in the galaxy NGC5907 (2XMM J151558.6+561810) was a PULX. It is the brightest 
PULX, with an X--ray luminosity $\sim 10^{41}$ erg\,s$^{-1}$, putting 
it in the category of the so--called hyperluminous sources  ($L_X\geq 10^{41}$ erg\,s$^{-1}$).

\subsubsection{M51 ULX8}
The ULX8 in the galaxy M51 was the first ULX system with a probable cyclotron feature: a
3.8$\sigma$ absorption line at 4.5 keV in its Chandra spectrum \citep{Brightman18}.
Since \citet{Middleton18} rule out the presence of a $10^{15}$\,G dipole magnetic field in the system,
it is reasonable to conclude that this line must be produced by electrons gyrating in a field
of strength $\lesssim 3\times 10^{11}$\,G.

\subsection{Transient PULXs}

\subsubsection{NGC300 ULX1}
This system in the galaxy NGC300 became active in X-rays and optical in 2010 and was initially 
classified as the supernova SN 2010da. After its unmasking as a supernova impostor,
it was initially thought to be a BeX system in a giant outburst. But its luminosity seems not 
to have varied by more than a factor 2 or 3 between 2010 and 2018 
\citep{Vasilopoulos18}, making 
this classification dubious. The nature of the companion is unclear \citep[see][and references therein]{Vasilopoulos18}. \citet{Carpanoetal18} found a
strong modulation in the 2016 X--ray data with a pulse period of 31.6 s. The X--ray signal shows spin--down glitches  \citep[also called `anti--glitches';][]{Rayetal18}.
If, as speculated by \citet{Binder16}, NGC 300 ULX1 were the first massive progenitor binary ever observed to evolve into an HMXB, this system would have to
be moved into the standard NSULX category.

\subsubsection{SMC X-3}
The Small Magellanic Cloud contains 121 High Mass X-ray Binaries (HMXBs) of which 64 
contain pulsars \citep[see][and references therein]{Townsend17}.
SMC X--3 is a well--known $\sim 7.8$s X-ray pulsar in a near--circular
orbit with a Be star.
In 2016 this system was observed to undergo a 5 month super--Eddington 
outburst \citep{WengX3,KoliopanosX3} during which the previously slowing pulsar was discovered to be rapidly spinning 
up. The observed spinup was 500 times larger than the previous spin--down. The angular 
momentum transferred by matter accreted during the five months giant outburst was greater than
that lost by magnetic braking over the preceding 18 yr \citep{Townsend17}. SMC 
X-3 is the only PULX where we clearly see the transition from an X--ray pulsar to a PULX and 
back, when first, on MJD 5168, it switched from spin-down mode to an extremely rapid 
spinup coincident with
super--Eddington luminosity, returning to the previous mode after five months \citep{Townsend17}.

\subsubsection{NGC 2403 ULX}
This transient X-ray pulsar in the nearby spiral galaxy NGC2403, known as 2XMM J073709.1+653544 (or CXOU J073709.1+653544), has a peak luminosity just above the $10^{39}\rm erg\,s^{-1}$ `limit', but its spinup
rate $> 10^{-10}\rm s^{-2}$ \citep{2007ApJ...663..487T} definitely puts it in the PULX category.

\subsubsection{NGC 1313 ULX}
This transient PULX  XMMU J031747.5-663010 in the isolated galaxy NGC3131 was observed once in 2004 by \citet{Trudolyubov08}. Its pulse period is much longer than observed in other PULX and
the pulse time--derivative has not been measured. This object should not be confused with sources NGC3131 X-1 or X-2 that are also ULXs \citep{Pintore12}.

\subsubsection{Swift J0243.6+6124}  
The first ULX candidate observed in the Galaxy, details of this presumed Be--X system 
depend on its uncertain distance \citep{Vandeneiden18}. However, its high spinup rate  
\citep{Doroshenko18} supports
ULX membership. It is the first known highly magnetized neutron--star jet emitter \citep{Vandeneiden18}.

\section{The PULX model of KLK17}

The only observables directly relating to PULX properties are the pulse period $P$, spinup 
$\dot\nu$ and luminosity $L$. In one other case 
a magnetic field is measured directly 
through a CRSF. No direct mass measurement exists, but 
for neutron stars the likely mass range is limited. 

PULXs are sharply distinct from other pulsing 
X--ray sources in their luminosities (by definition $L>10^{39}\rm erg\,s^{-1}$) and
their spinup rates ($\dot\nu > 10^{-10}\rm s^{-2}$), which are tightly correlated 
(Fig. \ref{fig:dotnul}). This correlation argues strongly that accretion provides the dominating 
torque in the system as assumed in e.g. KLK17 and \citet{Vasilopoulos18}, since
\begin{equation}
\dot\nu = \frac{\dot J(R_M)}{2\pi I} = \frac{\dot M (GMR_M)^{1/2}}{2\pi I} \propto \dot M^{6/7}
\label{eq:dotnudef}
\end{equation}
where $R_{\rm M} \propto \dot M^{-2/7}$ (e.g. Frank et al., 2002) 
is the magnetospheric radius and $I$ the neutron star's moment of inertia.

KLK17 used this fact to set up a model describing super--Eddington accretion on
to a magnetized neutron star. To describe the mass inflow they used the
 super--critical solution described in Section IV of \citet{SS73}  (not the celebrated thin--disc solution). In this picture, most of the super--Eddington mass supply is eventually expelled as
 a wind, which provides the beaming making ULXs appear super--Eddington.
 The accretor gains mass only at about its Eddington rate.  As mentioned in the 
 Introduction, systems such as Cyg X--2 give strong support to this picture.
 
 Within the
\textsl{spherization radius}, where the height of the radiation-pressure dominated accretion disc becomes comparable to the distance to the centre, the local disc luminosity is Eddington--limited, generating a strong outflow. 
Using $\dot m=\dot M/\dot M_{\rm Edd}$, where
\begin{align}
\dot M_{\rm Edd}  \equiv \frac{L_{\rm Edd}}{0.1c^2} & = 1.6\times 10^{18}m_1\,{\rm g\,s^{-1}} \nonumber \\
& =2.5\times 10^{-8}m_1\,\rm \msun\,yr^{-1},
\end{align}
the spherization radius is defined through the equation
\begin{equation}
R_{\rm sph}\simeq 1\times 10^6 \dot m_0 m_1\, \rm cm,
\label{eq:rsph}
\end{equation}
where $\dot M_0=\dot m_0\, \dot M_{\rm Edd}$  is the accretion rate at $R_{\rm sph}$, assumed equal to the mass-transfer rate. To keep the local energy release $GM\dot M(R)/R$
below $L_{\rm Edd}$ for ${R} < {R_{\rm sph}}$ we therefore have
\begin{equation}
\dot M(R) \simeq \dot m_0 \dot M_{\rm Edd}\frac{R}{R_{\rm sph}}.
\label{eq:mdotr}
\end{equation}

KLK17 assume that the \citet{SS73} model describes the accretion flow between $R_{\rm sph}$ and the magnetospheric radius $R_{\rm M}$ defined by
by the equations \citep[][]{FKR02} \begin{equation}
R_{\rm M} = 1.2 \times 10^8 q\, \dot m^{-2/7} m_1^{-3/7} \mu_{30}^{4/7}\, \rm cm,
\label{eq:rm}
\end{equation}
where $q$ is a factor taking into account the geometry of the
accretion flow at the magnetosphere. Then from Eq. (\ref{eq:mdotr})
\begin{equation}
\dot M(R_{\rm M}) \simeq {\dot M_0}\frac{R_{\rm M}}{R_{\rm sph}}.
\label{eq:mr}
\end{equation}
(see the unnumbered equation placed 22 lines from the top of p.353 of \citet{SS73}).

KLK17 found that for the first three PULXs known, ${R_{\rm M}}\approx {R_{\rm sph}}$, implying that the accretion flow inside the magnetosphere is highly super--Eddington 
when $\dot m_0 \gg 1$. As it is hypersonic but
forced to follow fieldlines it is highly dissipative, and one expects it to generate an outflow similar to 
that of \citet{SS73}, limiting the local luminosity to its Eddington value.

The total luminosity is then given by the unnumbered equation 11 lines from
the top of p. 353 of \citet{SS73}, which is equivalent to
\begin{equation}
L \simeq L_{\rm Edd}\left[1 + \ln \dot m_0\right].
\label{eq:Llog}
\end{equation}
In Sect. \ref{sec:discu} we discuss further the problems of modelling magnetospheric accretion flows 
in NSULXs\footnote{An equation deceptively formally similar to Eq. (\ref{eq:Llog}) holds 
in the physically completely distinct case of an advection dominated
accretion flow \cite[see eg.,][]{Poutanenetal07,Lasota16}.
But here $\dot M(R) = {\rm const}$ ({\it not} $\propto R$), and the binding energy which has not 
been radiated away would be still contained in the matter landing on the accretor's surface. For a 
black hole accretor,  Eq. (\ref{eq:Llog}) would correspond to the actual luminosity, but when the 
accretor is neutron star this is obviously not true, and Eq. (\ref{eq:Llog}) might be not a good 
approximation. We stress that, {\it by construction}, outflows from advection dominated flows must 
be weak. They cannot represent ULXs, where strong outflows are observed \citep{Pinto16}. 
{\citet{Mushtu19} propose a model with a mixture of outflow and advection but 
do not test it against the observed values of $\dot \nu$.}.
}
The luminosity 
from both parts of the super--Eddington outflow is expected to be beamed \citep[see, e.g.][and KLK17]{Kingetal01,King09,KL16,MK17}. Outside the magnetosphere the beaming factor is taken to be
\begin{equation}
b \simeq \frac{73}{\dot m^2}
\label{eq:b2}
\end{equation}
for the total beam solid angle $4\pi b$, valid for $\dot m_0 > 8.5$. \citet{King09} deduced from the $L_{\rm soft} \propto T^{-4}$ dependence observed
in a sample of ULXs by \citet{KP09}. Although some ULXs obey
$L_{\rm soft} \propto T^{4}$ 
\citep{Miller13}, these are probably black--hole systems with masses large enough to make
their luminosities sub--Eddington.

\citet{Mainieri10} found that the Local Group ULX luminosity function favours $b \sim \dot m^{-2}$. Also, as shown in \citet{KL16} and KLK17 and confirmed in the present paper, the approximate formula (\ref{eq:b2})
gives reasonable results when applied to PULXs,  and its main conclusion that these systems 
have $R_{M}\approx R_{\rm sph}$  is supported by observations \citep[see e.g.,][]{Walton18a}.

For accretion rates such that  radiation is geometrically beamed as described by Eqs. (\ref{eq:lbeam}) and  (\ref{eq:b}) one can deduce $\dot m_0$ from the observed X-ray luminosity $L=L_X$ by combining these two equations into
\begin{equation}
L_{X} \simeq 2.0 \times 10^{36}\dot m^2_0\left[1 + \ln \dot m_0\right]m_1.
\end{equation}
Having $\dot m_0$ we obtain $R_{\rm sph}$ from Eq. (\ref{eq:rsph}).

Then the second observed quantity, the observed spinup
\begin{equation}
\dot\nu=3.3 \times 10^{-11} q^{1/2}\dot m^{6/7}m_1^{6/7}\mu_{30}^{2/7}I_{45}^{-1}\rm s^{-2},
\label{eq:dotnumag}
\end{equation}
provides the values of  the magnetic moment $\mu$, which in turn allows us to calculate the values of $R_M$ and $\dot m(R_M)$ from  Eqs (\ref{eq:rm}) and  (\ref{eq:mr})\footnote{{{Here and elsewhere small numerical errors and inconsistencies in KLK17 have been corrected.}} }:
\begin{equation}
\mu_{30} = 0.04\,  q^{-7/4}\dot \nu^{3/2}_{-10} m_1^{1/2} I_{45}^{3/2}\, \rm G\,cm^{3}
\label{eq:mures}
\end{equation}
\begin{equation}
R_{\rm M} = 1.0 \times 10^7  \dot \nu^{2/3}_{-10} m_1^{1/3} I_{45}^{2/3}\, \rm cm
\label{eq:rmres}
\end{equation}
\begin{equation}
\dot m(R_{\rm M}) = 9.9\,  \dot \nu^{2/3}_{-10} m_1^{-2/3} I_{45}^{2/3},
\label{eq:mdotres}
\end{equation}
where $10^{-10}\dot \nu_{-10}=\dot \nu$.

The results are presented in Table \ref{tab:ulx3b}, with the values of the corotation radius
$R_{\rm co}\equiv \left(GMP_s^2/4\pi^2\right)^{1/3}$cm, the equilibrium period $P_{\rm eq} = 0.23 q^{7/6}m_1^{-1/3} \mu_{30}^{2/3}$\,s,
and the lower limit on the time to reach equilibrium at the present spinup rate
\begin{equation}
t_{\rm eq}\equiv \frac{1}{\dot \nu}\left(\frac{1}{P_{\rm eq}} - \frac{1}{P }\right).
\end{equation}

It is encouraging that for NGC300 ULX-1 the KLK17 model predicts a neutron--star magnetic 
field $B \approx 10^{12}$G, as 
the same value is deduced by \citet{Walton18} from the  CRSF observed in the X-ray 
spectrum of the PULX.  According to the KKL17
model all seven PULXs have magnetic fields in the range $10^{11}-10^{13}$G, i.e. below the 
value defining magnetars.

The predicted values of the beaming factor $b$ are in the range $\sim 0.06$ -- $0.6$, except for the hyperluminous source in NGC5709,
for which $b\approx 0.009$ \citep[similar to the values deduced by] [for HLX1 in ESO 243-49.]{KL14,Lasota18} For ULX1 in NGC300 we get $b\approx 0.18$, while \citet{Binder18} using our model obtain $b\approx 0.25$,
because to deduce $\dot m_0$ they use the average instead of the maximum luminosity. The beaming factor $b\approx 0.2$ for P13 in NGC7793 is consistent with observations of the X-ray irradiation of neutron-star companion
\citep{Motch18}.

We see that, as found by KLK17 for the three `standard' PULXs, all seven PULXs with 
known values of spinup rates $\dot \nu$
give $R_{\rm M}\sim R_{\rm sph}$, which as explained in the Introduction is probably the 
condition for observing pulses at all if mass transfer is super--Eddington.

Parametrizing  $R_{\rm M}=f R_{\rm sph}$, with $f\approx 0.3 - 1$, then from Eq. (\ref{eq:mr}) $m(R_{\rm M})= f\dot m_0$ (KLK17) and from Eqs 
(\ref{eq:dotnumag}) and (\ref{eq:mdotres}) one finds
\begin{equation}
\dot \nu = 7.8 \times 10^{-10}\, f^{7/6} q^{7/6} \mu^{2/3}_{30} m_1^{-1/7} I_{45}
^{-1},
\label{eq:nueq}
\end{equation}
in agreement with the observed spinup values of PULXs.

For M51 ULX8  -- not a PULX -- 
we know the value of the magnetic field. Assuming its X-ray radiation is beamed we deduce
$\dot m_0$ and so both $R_{\rm M}$ and $R_{\rm sph}$. 
{In this case $R_{\rm M}$ is slightly larger than $R_{\rm sph}$  (see Table \ref{tab:ulx3b}) but by no more than in some PULXs.
In view of the model uncertainties (see Section \ref{sec:discu})  we speculate that M51 ULX8 might
be a PULX and suggest that pulsations may be seen in future observations of this system. As pointed out by \citet{Brightman18}, longer
XMM-Newton observations would be needed to detect $\sim1$s pulsations if the pulsed fraction is $\lesssim 45\%$, as in most
PULXs. \citep[Interestingly the only exception is NGC 300 ULX1 in which a CRSF has been found by][]{Walton18}.}

\section{No magnetars in binary systems}

We have shown that the properties of PULXs are explicable with standard magnetic fields --
there is no need to invoke the presence of magnetars in them, and indeed the presence 
of magnetars is incompatible with the observations.

Our no--magnetar result
is consistent with the fact that no observed magnetar is a 
member of a binary system. We suggest here that there are good reasons for this absence.

All observed magnetic field strengths of neutron stars in X--ray binaries are in the range $10^8 - 10^{13}$G \citep[see, e.g.,][]{Revni16}. In contrast, {\sl all} $\sim 30$ known
magnetars (or ``candidate" magnetars) are isolated neutron stars \citep{Olausen14}. 

There is strong evidence that magnetars are (or even must be) {\it formed} in binary systems, 
but that the formation process usually leads to the
destruction of the binary \citep{Popov16}. Magnetar formation through neutron star mergers \citep[e.g.,][]{Giaco13,Piro17} obviously produces a single object.
Magnetars may be also produced in type Ibc supernova outbursts in massive binaries, as demonstrated by the magnetar CXOU J1647-45 in the young massive cluster
Westerlund 1 \citep{Clark14}. Such superluminous supernovae disrupt the  
binary and produce isolated magnetars. In principle, a small fraction of binaries could
survive the catastrophe: for example the very slow (pulse period $\sim 2.6$h) X-ray pulsar in the high-mass X-ray binary 4U 0114+65 could have a magnetar--strength magnetic
field \citep[][and references therein]{Sanjur17}. The problem with this channel as a model
for PULX formation is that strong neutron--star fields $\sim 10^{15}$G decay quickly ($\lesssim$ a 
few Myr) by several
orders of magnitude \citep{Mereghetti15}. \citet{Igoshev18} find that the age of 4U 0114+65 
is 2.4 -- 5 Myr since the formation of the neutron star, and discuss the
conditions for long survival (up to $10$ Myr) of a magnetar--strength field. One of 
these conditions is rapid cooling of the crust below $T \approx 3 \times 10^7$ K. This
is probably satisfied for slowly accreting neutron stars, but doubtful for
fast--rotating, Eddington--accreting NSULXs. One might try to escape these restrictions by 
postulating magnetar formation in hierarchical triple systems, but it is then hard to overcome
the twin problems that magnetar formation may well unbind the resulting loosely--bound daughter 
binary system, or that this system is too wide to come into contact within the field
decay time $\sim$ few Myr. It appears very unlikely that any, let alone {\sl all} NSULXs are 
survivors of binaries experiencing powerful type Ibc supernova explosions. Accordingly it is
unsurprising that our analysis does not suggest any candidate for a magnetar--strength field in
NSULXs.

\section{Discussion and conclusions}
\label{sec:discu}

We have extended the analysis of KLK17 to the current sample of PULXs and found that the 
condition $R_M \sim R_{\rm sph}$ for observable pulsing still holds
(in the language of \citet[][see their Fig. 4]{Middleton18} this corresponds to a ``critical" disc 
geometry). As in KLK17, 
we showed that this implies values of the spinup rate $\dot \nu > 10^{-10}\rm 
s^{-2}$, fully consistent with the observed PULX properties, and correlating strongly with their 
luminosities (see Figure \ref{fig:dotnul}).

Models invoking magnetar-strength magnetic fields \citep[e.g.,][]{Mushtu15} do not address
the observed of  PULXs spinup rates but as we have shown above, these spinup rates imply
normal-strength  ($\sim 10^{11} - 10^{13}$ G) fields.

Our calculations, like all other attempts to describe 
super--Eddington accretion flows, are based on a simple
approximate model, and minor discrepancies (in a few cases $R_M$ slightly exceeds
$R_{\rm sph}$) are to be expected. The magnetospheric radius is calculated from
formulae appropriate to thin--disc accretion, whereas we assume that the disc interacting with the 
magnetosphere is thick. The formula for the spherization radius
derived by \citet{SS73} corresponds to the ring where the disc height is equal to 
its radius, which might not be exact in reality \citep[see, e.g.][]{Poutanenetal07}.
Numerical simulations, although very impressive, cannot be of much help (even if they 
discuss the ``spherization radius"). Instead of an external 
thin disc, they have a torus \citep{Takahashi17,Takahashi18,David18}. The 
critical radii are uncertain by no more than factors $\sim 1$, so we expect (as here) the results
of KLK17 will remain robust in comparison with both observations and future numerical simulations.

We conclude that magnetar field strengths are not required to describe PULXs. This is reassuring because the presence of magnetars in binary systems, especially 
ultraluminous accreting systems, appears extremely unlikely.

\section*{Acknowledgments}
We are grateful to the anonymous referee for a very detailed and helpful report.
We thank Matt Middleton for useful advice and help.
This research was supported by the Polish NCN grant No. 2015/19/B/ST9/01099. 
ARK acknowledges partial support by the LABEX ``Institut Lagrange de Paris". 
The Theoretical Astrophysics Group at the University of Leicester is
supported by an STFC Consolidated Grant.
JPL acknowledges support from the French Space Agency CNES.

\appendix
\section{X-ray pulsars }
To illustrate the fact that  PULX differ from standard X-ray pulsars not only by the luminosity $> 10^{39}$\,erg/s  but also by their very large spinup rate $\dot \nu > 10^{-10}$\,s$^{-2}$,  we used the XRP sample with known luminosities and spin-up rates (Table A1) as compiled by \citet[][]{Ziolkowski85}.
\begin{center}
\begin{table}
{
\setlength{\tabcolsep}{1pt}
\caption{Observed properties of `classical' X-ray pulsars$^*$}
{\small
\hfill{}
\begin{tabular}{ ||l||c||l||c|} 
 \hline\hline
Name & $L_X (\rm max)$  [erg\,s$^{-1}$] & $P_s $ [s] &  $\dot \nu$ [s$^{-2}$]   \\
\hline\hline
SMC X-1 & $5.0 \times 10^{38}$  & 0.71 & $2.6 \times 10^{-11}$ \\
\hline
Cen X-3 & $5.0 \times 10^{37}$& 4.84 & $1.9 \times 10^{-12}$ \\
\hline
GX 1+4 & $ 4.0 \times 10^{37}$ & 122 & $5.2 \times 10^{-12}$\\
\hline 
4U 0115+63 &$ 3.4 \times 10^{37}$ & 3.61 & $2.2 \times 10^{-12}$ \\
\hline
A 0535+26 & $ 1.0 \times 10^{37}$  & 104 & $1.9 \times 10^{-12}$ \\
\hline
Her X-1 & $ 1.0 \times 10^{37}$ & 1.24 & $8.5 \times 10^{-14}$ \\
\hline
Vela X-1 & $ 1.5 \times 10^{36}$ & 283 & $5.6 \times 10^{-13}$ \\
\hline
GX 301-2 & $ 1.0 \times 10^{36}$ & 696 & $4.9 \times 10^{-13}$ \\
\hline\hline
$^*$ From \citet{Ziolkowski85}
\end{tabular}}
}
\hfill{}
\label{tb:XRP}\\
\vskip 0.2truecm 
\end{table}
\end{center}

\bsp

\label{lastpage}

\end{document}